\documentstyle[prb,aps,tighten,preprint]{revtex}

\begin{document}
\preprint{cond-mat/9406036}
\draft
\title{Conductance distribution of a quantum dot with non-ideal single-channel
leads}
\author{P. W. Brouwer and C. W. J. Beenakker}
\address{Instituut-Lorentz, University of Leiden, P.O. Box 9506, 2300 RA
Leiden, The Netherlands}
\date{Submitted to Physical Review B}
\maketitle

\begin{abstract}
The entire distribution is computed of the conductance of a quantum dot
connected to two electron reservoirs by leads with a single propagating mode,
for arbitrary transmission probability $\Gamma$ of the mode. The theory bridges
the gap between previous work on ballistic leads ($\Gamma = 1$) and on
tunneling point contacts ($\Gamma \ll 1$). \bigskip\\
%PACS numbers
\pacs{PACS numbers: 05.45.+b, 73.40.Gk, 72.10.Bg, 72.15.Rn}
% 05.      Statistical physics and thermodynamics
% 05.45.+b Theory and models of chaotic systems
% 73.      Electronic structure and electrical properties of surfaces,
%          interfaces, and thin films
% 73.40.Gk Tunneling
% 72.      Electronic transport in condensed matter
% 72.10.Bg General formulation of transport theory
% 72.15.Rn Quantum localization
\end{abstract}

An ensemble of mesoscopic systems has large sample-to-sample fluctuations in
its transport properties, so that the average is not sufficient to characterize
a single sample. To determine the complete distribution of the conductance is
therefore a fundamental problem in this field. Early work focused on an
ensemble of disordered wires. (See Ref.\ \ref{AltshulerLeeWebb} for a review).
The distribution of the conductance in that case is either normal or
log-normal, depending on whether the wires are in the metallic or insulating
regime. Recently, it was found that a ``quantum dot'' has a qualitatively
different conductance
distribution.\cite{PrigodinEfetovIida,BarangerMello,JalabertPichardBeenakker}
A quantum dot is a small confined region, having a large level spacing compared
to the thermal energy, which is weakly coupled by point contacts to two
electron reservoirs. The classical motion within the dot is assumed to be
ballistic and chaotic. An ensemble consists of dots with small variations in
shape or in Fermi energy. The capacitance of a dot is assumed to be
sufficiently large that the Coulomb blockade can be ignored, i.e. the electrons
are assumed to be non-interacting. Two altogether different approaches have
been taken to this problem.

Baranger and Mello\cite{BarangerMello}, and Jalabert, Pichard, and one of the
authors\cite{JalabertPichardBeenakker} started from random-matrix
theory.\cite{Mehta} The scattering matrix $S$ of the quantum dot was assumed to
be a member of the circular ensemble of $N \times N$ unitary matrices, as is
appropriate for a chaotic billiard.\cite{BlumelSmilansky,Smilansky} In the
single-channel case ($N=1$), the distribution $P(T)$ of the transmission
probability $T$ (and hence of the conductance $G = (2e^2/h)T$) was found to be
\begin{equation}
  P(T) = \case{1}{2} \beta T^{-1 + \beta/2} \label{BallisticDistribution},
\end{equation}
where $\beta \in \{1,2,4\}$ is the symmetry index of the ensemble ($\beta = 1$
or $2$ in the absence or presence of a time-reversal-symmetry breaking magnetic
field; $\beta = 4$ in zero magnetic field with strong spin-orbit interaction).
Eq.\ (\ref{BallisticDistribution}) was found to be in good agreement with
numerical simulations of transmission through a chaotic billiard connected to
ideal leads having a single propagating mode.\cite{JalabertPichardBeenakker}
(The case $\beta=4$ was not considered in Ref.\
\ref{JalabertPichardBeenakker}.)

Previously, Prigodin, Efetov, and Iida\cite{PrigodinEfetovIida} had applied the
method of supersymmetry to the same problem, but with a different model for the
point contacts. They considered the case of broken time-reversal symmetry
($\beta=2$), for which Eq.\ (\ref{BallisticDistribution}) would predict a
uniform conductance distribution. Instead, the distribution of Ref.\
\ref{PrigodinEfetovIida} is strongly peaked near zero conductance. The tail of
the distribution (towards unit transmission) is governed by resonant tunneling,
and is consistent with earlier work by Jalabert, Stone, and
Alhassid\cite{JalabertStoneAlhassid} on resonant tunneling in the
Coulomb-blockade regime.

It is the purpose of the present paper to bridge the gap between these two
theories, by considering a more general model for the coupling of the quantum
dot to the reservoirs. Instead of assuming ideal leads, as in Refs.\
\ref{BarangerMello} and \ref{JalabertPichardBeenakker}, we allow for an
arbitrary transmission probability $\Gamma$ of the propagating mode in the
lead, as a model for coupling via a quantum point contact with conductance
below $2e^2/h$. Eq.\ (\ref{BallisticDistribution}) corresponds to $\Gamma=1$
(ballistic point contact). In the limit $\Gamma \ll 1$ (tunneling point
contact) we recover, for $\beta = 2$, the result of Ref.\
\ref{PrigodinEfetovIida}. We consider also $\beta = 1$ and $4$ and show that
--- in contrast to Eq.\ (\ref{BallisticDistribution}) --- the limit $\Gamma \ll
1$ depends only weakly on the symmetry index $\beta$. In the crossover region
from ballistic to tunneling conduction we find a remarkable $\Gamma$-dependence
of the conductance fluctuations: The variance is monotonically decreasing for
$\beta = 1$ and $2$, but it has a {\em maximum} for $\beta = 4$ at $\Gamma =
0.74$.

The system under consideration is illustrated in the inset of Fig.\
\ref{fig1}b. It consists of a quantum dot with two single-channel leads
containing a tunnel barrier (transmission probability $\Gamma$). We assume
identical leads for simplicity. The transmission properties of this system are
studied in a transfer matrix formulation. The transfer matrix $M_d$ of the
quantum dot can be parameterized as
\cite{MelloPichard,StoneMelloMuttalibPichard}
\begin{equation}
  M_d = \left( \begin{array}{cc} u_1 & 0 \\ 0 & v_1 \end{array} \right)
        \left( \begin{array}{cc} \sqrt{1+\lambda_d} & \sqrt{\lambda_d} \\
                       \sqrt{\lambda_d} & \sqrt{1+\lambda_d} \end{array}
\right)
        \left( \begin{array}{cc} u_2 & 0 \\ 0 & v_2 \end{array} \right),
  \label{Mparam}
\end{equation}
where the parameter $\lambda_d$ is related to the transmission probability
$T_d$ of the dot by
\begin{equation}
  T_d = (1 + \lambda_d)^{-1}.
\end{equation}
The numbers $u_j$ and $v_j$ satisfy constraints that depend on the symmetry
of the Hamiltonian of the quantum dot:
\begin{equation}
  \begin{array}{lclclcl}
  u_j &=& e^{ i \phi_j} a_j, & \ &
  v_j &=& e^{-i \phi_j} a_j,
  \label{uvparam}
  \end{array}
\end{equation}
with $a_j$ a real ($\beta = 1$), complex ($\beta = 2$), or real quaternion
($\beta = 4$) number of modulus one. In general the choice for $u_j$ and $v_j$
and their parameterisation (\ref{uvparam}) is not unique. Uniqueness can be
achieved by requiring that
\begin{equation}
  \begin{array}{lcl}
    a_1 = 1, & \ & 0 \le \phi_j < \pi\ \ (j=1,2).
  \end{array}
\end{equation}

As in Refs.\ \ref{BarangerMello} and \ref{JalabertPichardBeenakker}, we assume
that the scattering matrix $S_d$ of the quantum dot is a member of the circular
ensemble, which means that $S_d$ is uniformly distributed in the unitary group
(or the subgroup required by time reversal and/or spin rotation symmetry). The
corresponding probability distribution of the transfer matrix $M_d$ is
\begin{equation}
  P_d(M_d)\, dM_d = \case{1}{2} \beta (1 + \lambda_d)^{-1-\beta/2}
    d\lambda_d d\phi_1 d\phi_2 da_2.
\end{equation}
The transfer matrix $M_b$ of the tunnel barrier in the lead is given by
\begin{equation}
  M_b = \left( \begin{array}{cc} \sqrt{1+\mu} & \sqrt{\mu_{\vphantom{d}}} \\
               \sqrt{\mu_{\vphantom{d}}} & \sqrt{1+\mu} \end{array} \right),
\end{equation}
with $\mu = (1 + \Gamma)^{-1}$.
The transfer matrix $M$ of the total system follows from the matrix product
\begin{equation}
  M = M_b M_{d} M_b.
  \label{Mmult}
\end{equation}

{}From Eqs.\ (\ref{Mparam})--(\ref{Mmult}) we straightforwardly compute the
transmission probability $T$ of the total system and its probability
distribution $P(T)$. The result for $T$ is
\begin{eqnarray}
  T &=& \left(  \vphantom{\sqrt{\lambda_d}}
                 1+ \lambda_d + m \lambda_d \cos^2\psi_- +
                 m(\lambda_d+1) \cos^2\psi_+ + \nonumber \right. \\ &&\
                 \left. 2\sqrt{\lambda_d(\lambda_d+1)m(m+1)}
                   \cos\psi_- \cos\psi_+ \right)^{-1},
  \label{lambdarel}
\end{eqnarray}
where we have abbreviated
\begin{equation}
  \begin{array}{cc}
  m = 4(1-\Gamma)\Gamma^{-2}, & \psi_\pm = \phi_1 \pm \phi_2.
  \end{array}
\end{equation}
The variables $a_j$, and with them all $\beta$-dependence, drop out of this
expression. Eq.\ (\ref{lambdarel}) can be inverted\cite{Inverting} to yield
$\lambda_d$ in terms of $\phi_1$ and $\phi_2$ for given $T$ and $\Gamma$. The
probability distribution $P(T)$ then follows from
\begin{equation}
  P(T) = {\beta \over 2 \pi^2} \int_0^{\pi} d\phi_1 \int_0^{\pi} d\phi_2 \
      (1 + \lambda_d)^{-1-\beta/2}
   \left|
     {\partial \lambda_d \over \partial T}
   \right|,
  \label{integral}
\end{equation}
where the integration is over all $\phi_i \in (0,\pi)$ for which $\lambda_d$ is
real and positive.

For $\Gamma = 1$ the function $P(T)$ is given by Eq.\
(\ref{BallisticDistribution}), as found in Refs.\ \ref{BarangerMello} and
\ref{JalabertPichardBeenakker}. In Fig.\ \ref{fig1} the crossover from a
ballistic to a tunneling point contact is shown. For $\Gamma \ll 1$ and $T \ll
1$, $\Gamma^2 P(T)$ becomes a $\Gamma$-independent function of $T/\Gamma^2$,
which is shown in the inset of Fig. \ref{fig1}c. Several asymptotic expressions
for $P(T)$ can be obtained from Eq.\ (\ref{integral}) for $\Gamma \ll 1$,
\begin{mathletters}
\label{scaleres}
\begin{eqnarray}
\beta = 1:\ P(T) =& \displaystyle \lefteqn{
  \left\{ \begin{array}{l} \displaystyle
           {8 \over\pi^2 \Gamma_{\vphantom{M}}} T^{-1/2} \\
           \displaystyle {\Gamma^{\vphantom{M}} \over \pi^2} T^{-3/2}
          \end{array} \right.}
          \hphantom{24 T\Gamma {3\Gamma^4 + 4 T\Gamma^{2} + 3 T^2
           \over (\Gamma^2 + 4T)^{9/2}}}
    & \begin{array}{l}
      \vphantom{{8 \over\pi^2 \Gamma_M}} \ (T \ll \Gamma^2), \\
      \vphantom{{\Gamma^M \over \pi^2}}  \ (\Gamma^2 \ll T \ll 1),
  \end{array}
  \label{scaleres1} \\
  \beta = 2:\ P(T) =&
    \lefteqn{4\Gamma {\Gamma^2 + T \over (\Gamma^2 + 4T)^{5/2}}}
    \hphantom{\displaystyle
           24 T\Gamma {3\Gamma^4 + 4 T\Gamma^{2} + 3 T^2
           \over (\Gamma^2 + 4T)^{9/2}}} & \
   \begin{array}{l} (T \ll 1), \end{array}
  \label{scaleres2} \\
  \beta = 4:\ P(T) =&
    \lefteqn{24 T\Gamma {3\Gamma^4 + 4 T\Gamma^{2} + 3 T^2
           \over (\Gamma^2 + 4T)^{9/2}}}
    \hphantom{\displaystyle
           24 T\Gamma {3\Gamma^4 + 4 T\Gamma^{2} + 3 T^2
           \over (\Gamma^2 + 4T)^{9/2}}} & \
  \begin{array}{l} (T \ll 1). \end{array} \label{scaleres3}
\end{eqnarray}
\end{mathletters}%
The $\beta = 2$ expression (\ref{scaleres2}) for $P(T)$ in the tunneling regime
agrees precisely with the supersymmetry calculation of Prigodin, Efetov, and
Iida.\cite{PrigodinEfetovIida,Compare} Eq.\ (\ref{scaleres}) does not cover the
range near unit transmission. As $T \rightarrow 1$ (and $\Gamma \ll 1$), $P(T)
\rightarrow c_\beta \Gamma$, with $c_1 = \case{1}{2\pi}$, $c_2 = \case{1}{4}$,
and $c_4 = \case{3}{8}$.

A quite remarkable feature of the quantum dot with ideal leads is the strong
$\beta$-dependence of $P(T)$ (cf.\ Fig.\ \ref{fig1}a). For $\Gamma \ll 1$, the
$\beta$-dependence is much less pronounced. For $T \gg \Gamma^2$ the leads
dominate the transmission properties of the total system, thereby suppressing
the $\beta$-dependence of $P(T)$ (although not completely). For very small
transmission coefficients ($T \ll \Gamma^2$) the non-ideality of the leads is
of less importance, and the characteristic $\beta$-dependence of Eq.\
(\ref{BallisticDistribution}) is recovered (see inset of Fig.\ \ref{fig1}c).

The moments of $P(T)$ can be computed in closed form for all $\Gamma$ directly
from Eq.\ (\ref{lambdarel}). The first two moments are (recall that $m =
4(1-\Gamma)\Gamma^{-2}$):
\begin{eqnarray}
  \langle T \rangle &=& \left\{ \begin{array}{lcl}
  \lefteqn{\case{1}{2} m^{-1} \left[ \sqrt{1+m} - \case{1}{\sqrt{m}}
  \ln(\sqrt{1+m} + \sqrt{m}) \right]}
  \hphantom{\case{3}{64} m^{-2} \left[ \left( 4 m - 18 \right) \sqrt{1+m} +
    \left({18 \over \sqrt{m}} + 8 \sqrt{m} \right) \ln(\sqrt{1+m} + \sqrt{m})
  \right]}
  & \ & (\beta = 1), \\
  \lefteqn{\case{2}{ 3} m^{-2} \left[(          m - 2 ) \sqrt{1+m} +  2
\right]}
  & \ & (\beta = 2), \\
  \lefteqn{\case{4}{15} m^{-3} \left[(3 m^2 - 4 m + 8 ) \sqrt{1+m} - 32
\right]}
  & \ & (\beta = 4),
   \end{array} \right. \\
  \langle T^2 \rangle &=& \left\{ \begin{array}{lcl}
  \case{3}{64} m^{-2} \left[ \left( 4 m - 18 \right) \sqrt{1+m} +
    \left({18 \over \sqrt{m}} + 8 \sqrt{m} \right) \ln(\sqrt{1+m} + \sqrt{m})
  \right] & \ & (\beta = 1), \\
  \case{4}{15} m^{-3}\left[ \left(m^2 + 2 m + 16 \right) \sqrt{1+m} - 10 m - 16
  \right] & \ & (\beta = 2), \\
  \case{4}{35} m^{-4}\left[ \left(3 m^3 + 2 m^2 - 40 m -144 \right) \sqrt{1+m}
    + 112 m + 144 \right] & \ & (\beta = 4).
  \end{array} \right.
\end{eqnarray}
For $\Gamma \ll 1$ one has asymptotically
\begin{equation}
  \langle T^n \rangle = {\beta \Gamma \over 2(\beta + 1)} \prod_{j=1}^{n-1}
   {(\beta + 2j)(2j-1) \over 2j(\beta + 2j + 1)}.
\end{equation}
The $\Gamma$-dependence of the variance $\mbox{Var}\ T = \langle T^2 \rangle -
\langle T \rangle^2$ of the transmission probability is shown in Fig.\
\ref{fig2}. In the crossover regime between a ballistic point contact ($\Gamma
= 1$) and a tunneling point contact ($\Gamma \ll 1$), the three symmetry
classes show striking differences. For $\beta = 1$ and $2$ the conductance
fluctuations decrease monotonically upon decreasing $\Gamma$, whereas they show
non-monotonic behavior for $\beta = 4$. Notice also that the transition $\beta
= 1 \ \rightarrow \ \beta = 2$, by application of a magnetic field, reduces
fluctuations for $\Gamma > \Gamma_c$ but increases fluctuations for $\Gamma <
\Gamma_c$, where $\Gamma_c = 0.92$.

In summary, we have computed the transmission probability of a ballistic and
chaotic cavity for all possible values of the symmetry index $\beta$ and for
arbitrary values of the transparency $\Gamma$ of the single-channel leads. Our
results describe the conductance of a quantum dot in the crossover regime from
a coupling to the reservoirs by ballistic to tunneling point contacts. The
theory unifies and extends known
results.\cite{PrigodinEfetovIida,BarangerMello,JalabertPichardBeenakker}
The characteristic $\beta$-dependence of the distribution function that was
found for ideal leads [Eq.\ (\ref{BallisticDistribution})] is strongly
suppressed for transmission probabilities $T$ larger than $\Gamma^2$. A closely
related phenomenon is the non-trivial $\Gamma$-dependence of the conductance
fluctuations for the three symmetry classes. The theory is relevant for
experiments on chaotic scattering in quantum dots with adjustable point
contacts, which are of great current interest.\cite{MRWHG,KMMPS,CBPW}

This work was supported by the Dutch Science Foundation NWO/FOM.

\begin{figure}
\caption{Distribution of the transmission probability $T$ through a quantum dot
with non-ideal single-channel leads, for three values of the transmission
probability $\Gamma$ of the leads. The curves are computed from Eq.\
(\protect\ref{integral}) for each symmetry class ($\beta=1,2,4$). The inset of
(b) shows the quantum dot, the inset of (c) shows the asymptotic behavior of
$P(T)$ for $\Gamma \ll 1$ on a log-log scale.\label{fig1}}
\end{figure}

\begin{figure}
\caption{Variance of the transmission probability $T$ as a function of the
transmission probability of the leads $\Gamma$.\label{fig2}}
\end{figure}


\begin{references}
\bibitem{Inverting} Inversion of Eq.\ (\ref{lambdarel}) requires some care.
Since a shift $\psi_+ \rightarrow \psi_+ + \pi$ changes the sign of the term
containing the square root in Eq.\ (\ref{lambdarel}), solving for $\lambda_d$
with $\psi_1$ and $\psi_1 + \pi$ yields exactly two (complex) solutions in
total. This allows one to construct a single-valued function
$\lambda_d(\psi_+,\psi_-)$ such that these two solutions are given by
$\lambda_d(\psi_+,\psi_-)$ and $\lambda_d(\psi_+ + \pi,\psi_-)$. This function
$\lambda_d$ is understood as the inverse of Eq.\ (\ref{lambdarel}).
\bibitem{Compare} To compare with Ref.\ \ref{PrigodinEfetovIida} we identify
$\alpha_1 = \alpha_2 = {1 \over 2} \Gamma \ll 1$ and take the limit $\alpha
\rightarrow 0$ of Eq.\ (7) in that paper. This yields our Eq.\
(\ref{scaleres2}). Here $\alpha_1$, $\alpha_2$, and $\alpha$ are, respectively,
the level broadening (divided by the level spacing) due to coupling to lead 1
and 2, and due to inelastic scattering processes (which we have not included in
our formulation, whence $\alpha \rightarrow 0$).
\end{references}
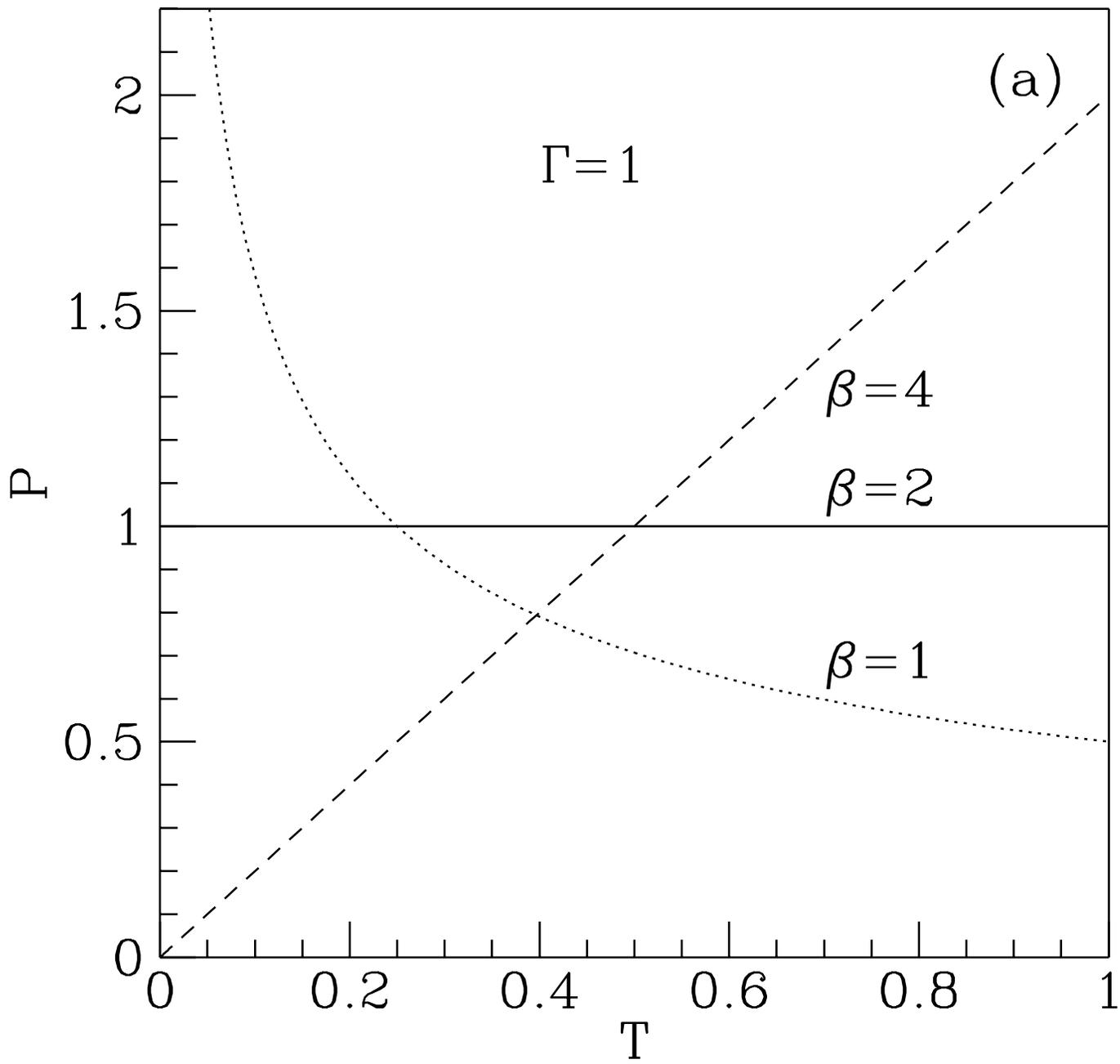
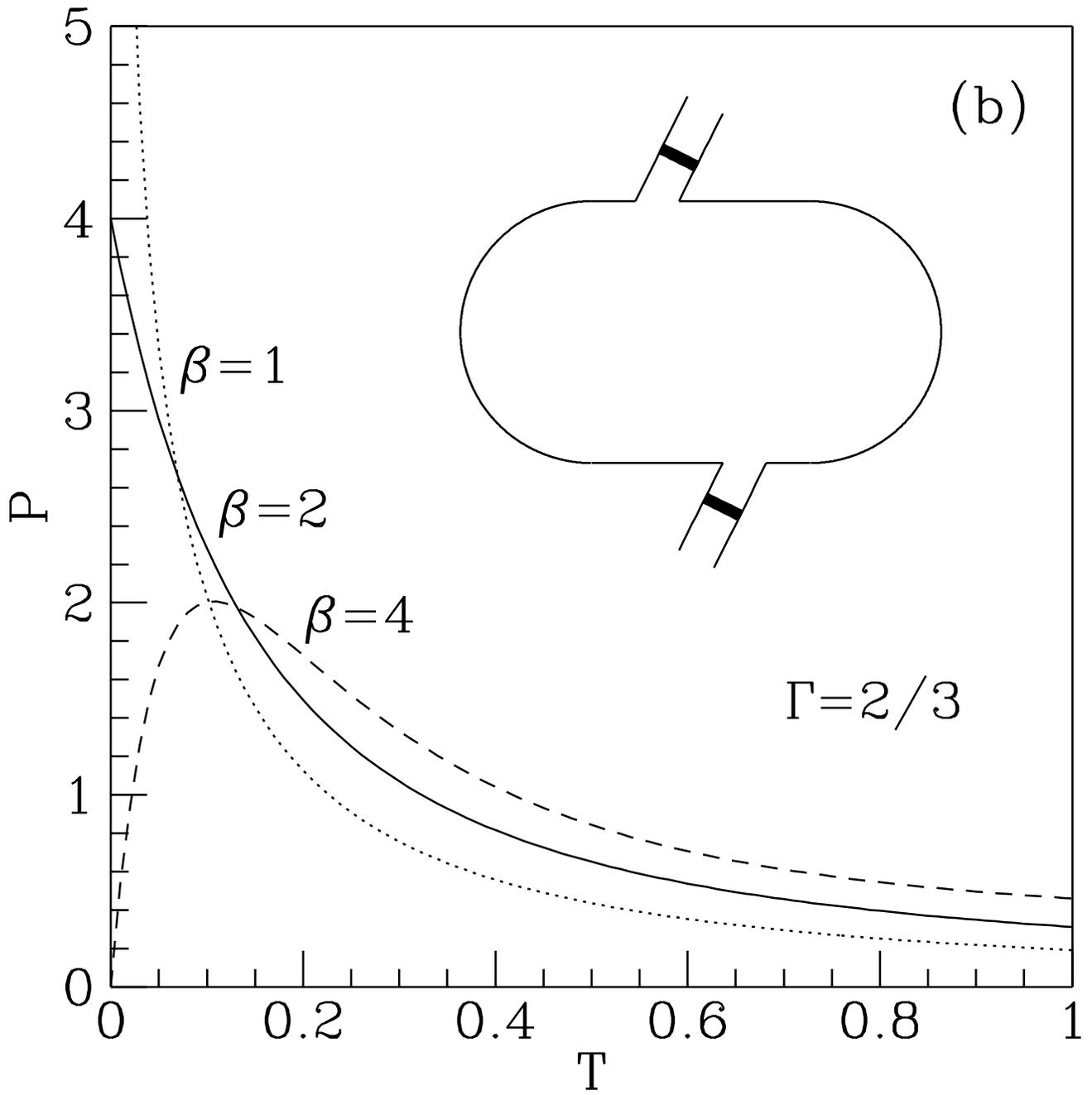
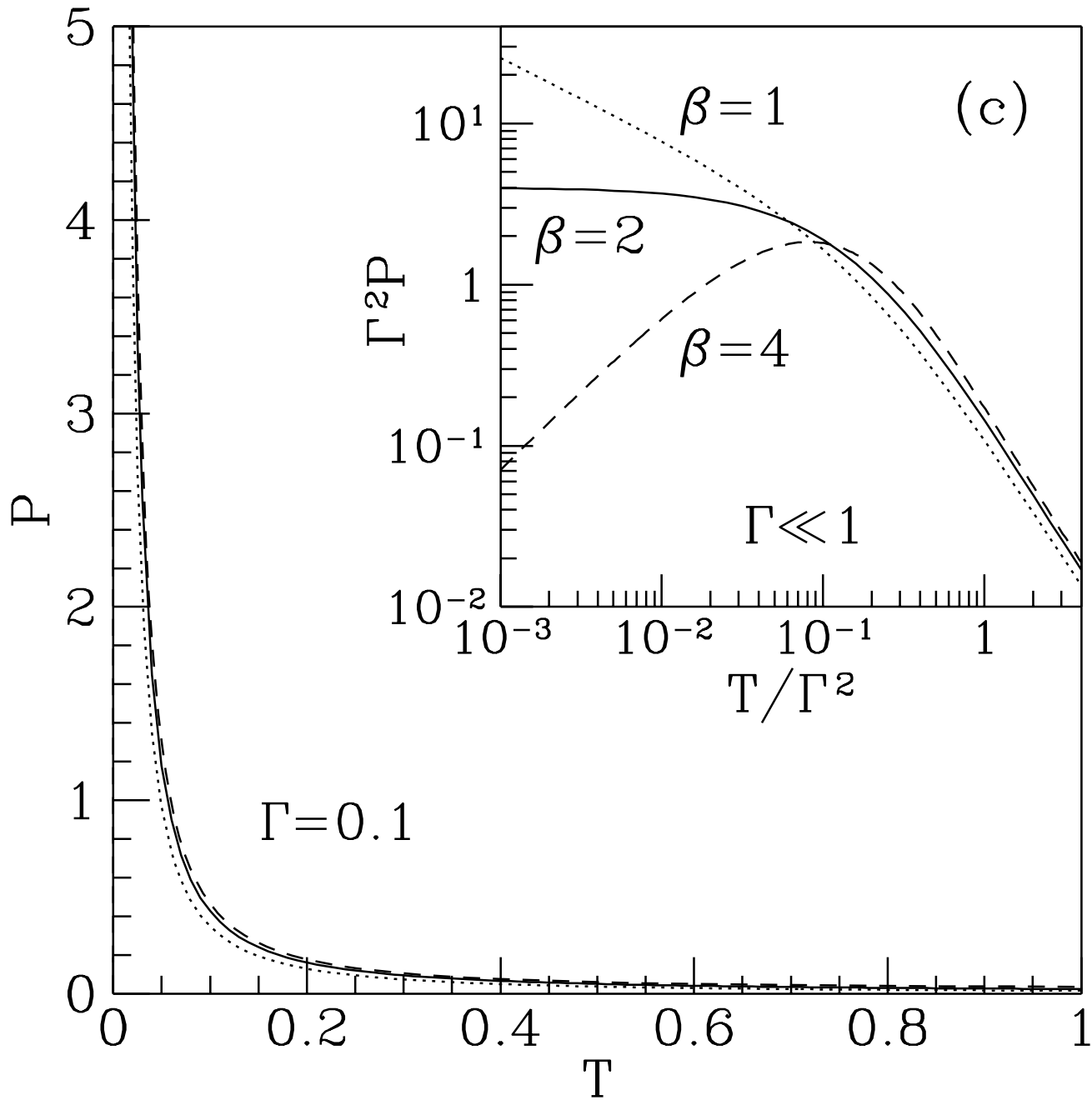
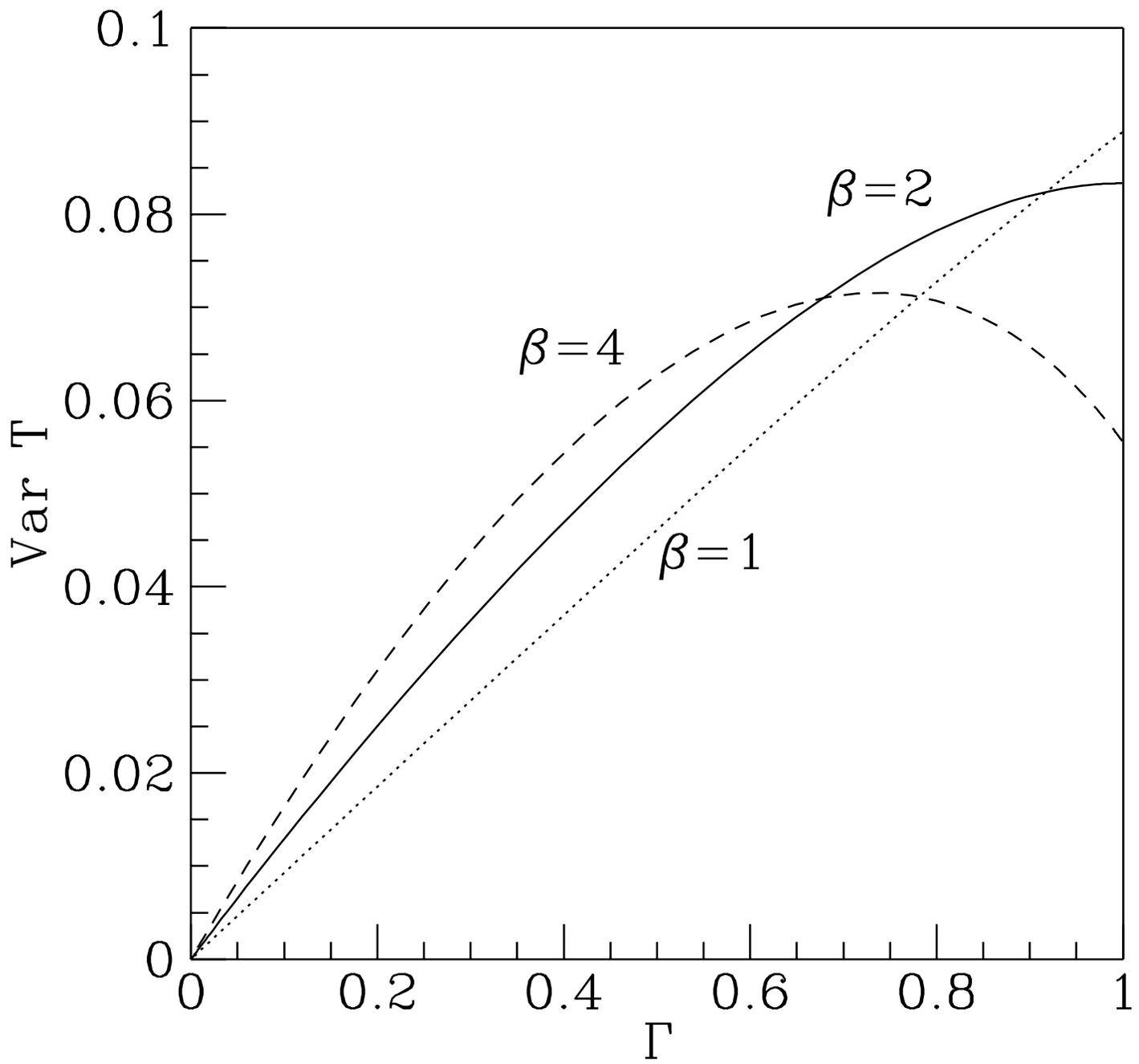
\end{document}